\documentclass[%
  journal = jpclcd,%
  layout = twocolumn,%
  ]{achemso}
\usepackage[latin1]{inputenc}
\usepackage{graphicx}
\usepackage{amsmath,amssymb}
\usepackage{amsthm}
\usepackage{float}
\usepackage[version=3]{mhchem}
\usepackage{booktabs}
\usepackage{widetable}
\usepackage{setspace}
\usepackage{multirow}
\usepackage{upgreek}
\usepackage{physics}
\usepackage{braket}
\usepackage{cleveref}
\usepackage{longtable}

\usepackage{braket}

\usepackage{threeparttable}
\usepackage[varg]{txfonts}
\usepackage[font=footnotesize,labelfont=bf]{caption}
\usepackage[small]{titlesec}
\setkeys{acs}{articletitle = true}
\setkeys{acs}{maxauthors=15}
\setkeys{acs}{etalmode=truncate}
\usepackage{etoolbox}
\makeatletter
\patchcmd{\ttlh@hang}{\parindent\z@}{\parindent\z@\leavevmode}{}{}
\patchcmd{\ttlh@hang}{\noindent}{}{}{}
\makeatother

\titleformat{\subsubsection}[runin]
{\normalfont\footnotesize\bfseries}{\thesubsubsection}{0.5em}{}
\titlespacing{\subsubsection}{0.0em}{0.0em}{0.5em}

\crefname{equation}{equation}{Equations}
\Crefname{equation}{Equation}{Equations}
\crefrangelabelformat{equation}{(#3#1#4--#5#2#6)}

\crefmultiformat{equation}{equations (#2#1#3}{, #2#1#3)}{#2#1#3}{#2#1#3}
\Crefmultiformat{equation}{Equations (#2#1#3}{, #2#1#3)}{#2#1#3}{#2#1#3}

\usepackage{xcolor}
\definecolor{aaltoOrange}{RGB}{255,121,0}%
\definecolor{aaltoBlue}{RGB}{0,101,189}%

\newcommand{\onlinecite}[1]{\hspace{-1 ex} \nocite{#1}\citenum{#1}}

\title{\Large{A balanced treatment of static and dynamic correlation in free- and Mg-porphyrin}}

\author {\normalsize{Marc Dvorak}}
\affiliation{\small{Department of Applied Physics, Aalto University School of Science, 00076-Aalto, Finland}}
\email{marc.dvorak@aalto.fi}
\author {\normalsize{Patrick Rinke}}
\affiliation{\small{Department of Applied Physics, Aalto University School of Science, 00076-Aalto, Finland}}
\let\oldmaketitle\maketitle
\let\maketitle\relax


\begin{tocentry}
\centering
\includegraphics[width=\columnwidth]{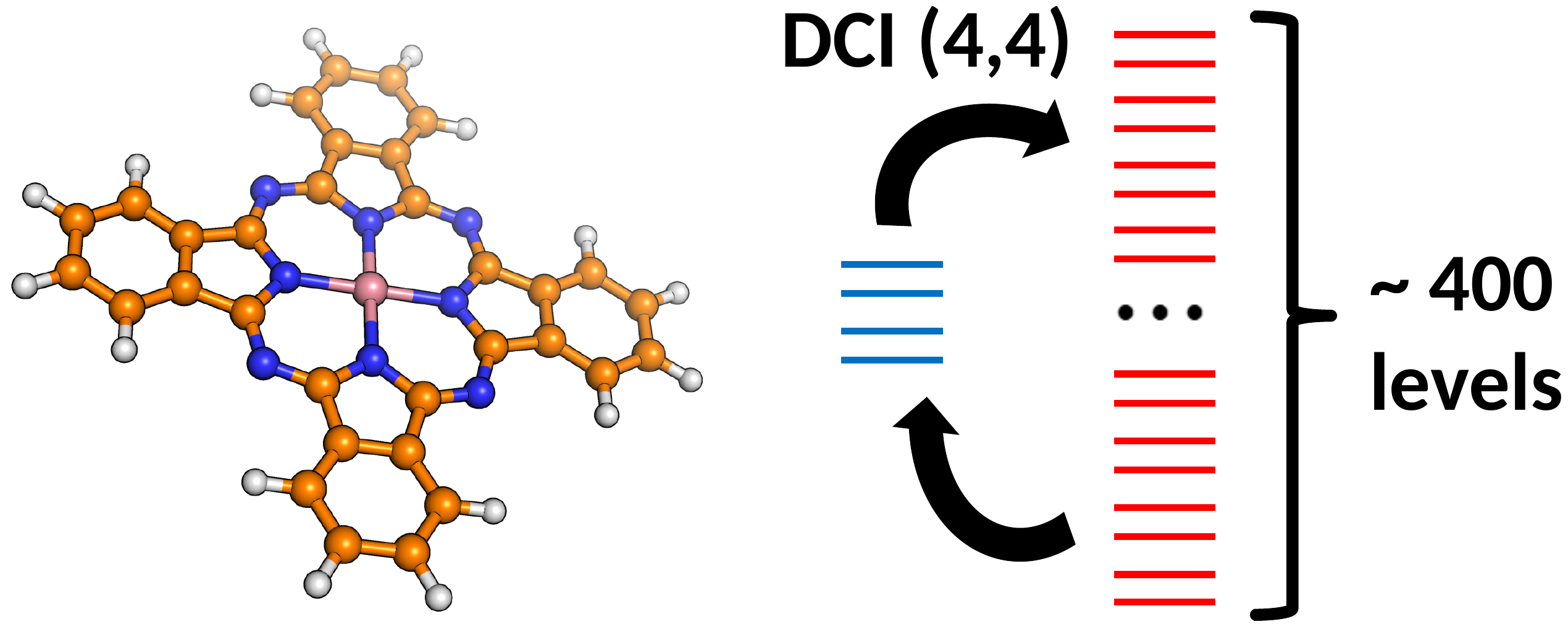}
\end{tocentry}

\begin{document}
\linespread{1.1}
\fontsize{10}{12}\selectfont
\twocolumn[
  \begin{@twocolumnfalse}
    \oldmaketitle
    \begin{abstract}
\fontsize{10}{12}\selectfont  
We present an \textit{ab-initio} dynamical configuration interaction (DCI) study of free- and Mg-porphyrin. DCI is a recently developed active space theory based on the L\"owdin downfolding technique. In the active space, static correlation is described exactly with full configuration interaction. In the high energy, dynamically correlated subspace, we treat correlation at the quasiparticle level in the $GW$ approximation of Green's function theory. The final theory combines wave function and Green's function methods to give a balanced description of static and dynamic correlation. The theory and algorithm give a multireference treatment of ground and excited states for low computational cost. The four orbital Gouterman model of porphyrin offers an ideal active space in a large, correlated system to test the cost and accuracy of the embedding for large systems. Our parameter free, fully \textit{ab-initio} DCI calculations in the minimal four-level active space agree well with both experiment and more expensive benchmark theories for the $Q_x$ and $Q_y$ transitions of free- and Mg-porphyrin. Studying the convergence of the excitation energies suggests that DCI correctly captures size extensive correlation effects, making it a promising active space theory for large, strongly-correlated systems.
    \end{abstract}
  \end{@twocolumnfalse}
  ]

A longstanding challenge in theoretical chemistry is the accurate description of systems with static and dynamic correlation. These two ubiquitous terms can be explained in many different ways. From a spectroscopic point of view, for example, sharp peaks in optical spectra can be assigned to a single molecular transition. These renormalized excitations mostly retain the character of their underlying bare excitation and can be considered dynamically correlated. In contrast, merged peaks which overlap with each other indicate several eigenstates close in energy which mix when the system is probed. Such an excitation with many strongly contributing eigenstates can be considered statically correlated.

\begin{figure}[htb]
\begin{centering}
\includegraphics[width=0.6\columnwidth]{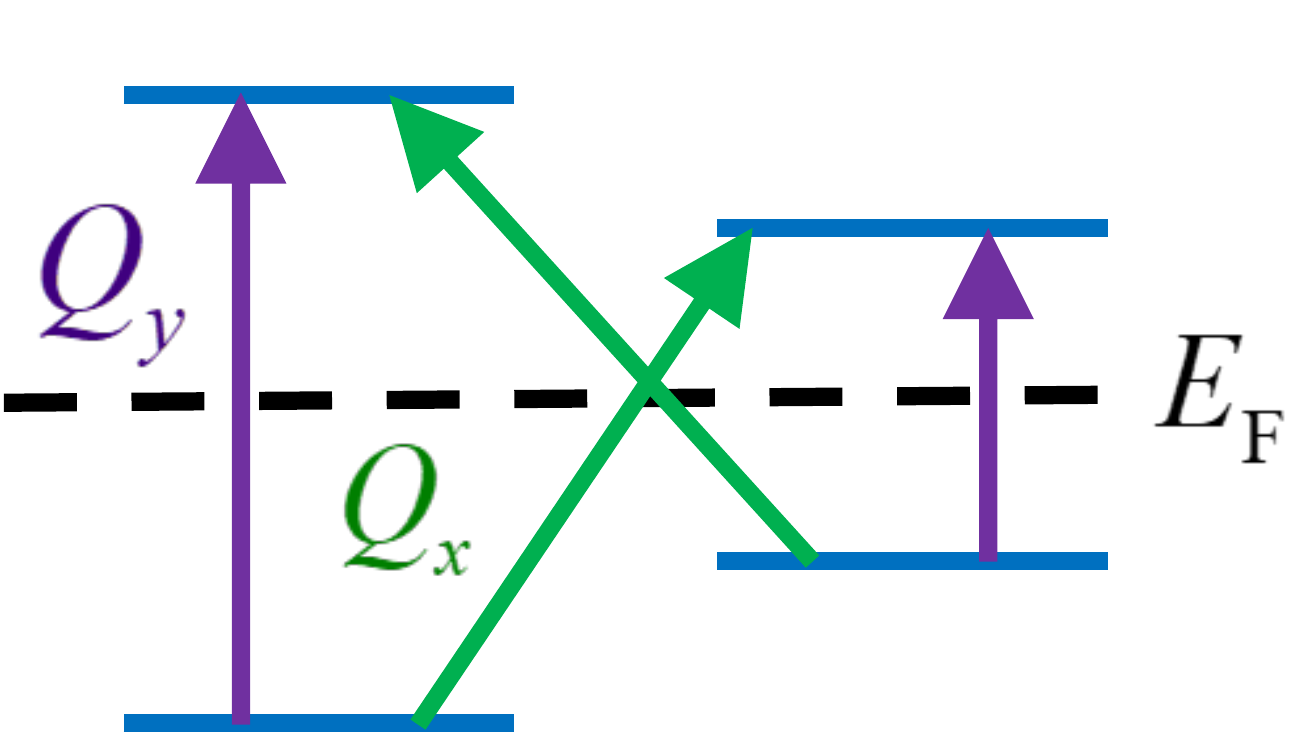}
\caption{Mixing of frontier molecular orbitals forming the $Q_x$ and $Q_y$ transitions in the four-level Gouterman model. In Mg-porphyrin, the $Q_x$ and $Q_y$ transitions merge into a degenerate $Q$ excitation. \label{gouterman_model} }
\end{centering}
\end{figure}

Many real systems contain both types of correlation. While it can be difficult to know \textit{ab-initio} if this is the case, chemical intuition and partitioning the physical problem can help. Systems with both types of correlation are suited for active space (AS) or quantum embedding theories in which a small portion of the problem is treated with high accuracy, and the remaining degrees of freedom are treated with a less expensive theory.\cite{jacs_leighton} Porphyrins are one such system. Their minimal four-level Gouterman model, shown in Fig.~\ref{gouterman_model}, qualitatively describes low-energy physics.\cite{jms_gouterman}  Figure~\ref{gouterman_model} illustrates how the four Gouterman orbitals contribute to the $Q_x$ and $Q_y$ optical excitations. In the higher symmetry Mg-porphyrin, $Q_x$ and $Q_y$ merge into a degenerate $Q$ excitation (not shown in the figure). These four statically correlated orbitals are embedded in and dynamically correlated with a large set of orbitals hosted by the ligands. To go beyond the qualitative Gouterman model and reach quantitative accuracy, dynamic correlation from these surrounding orbitals is essential.

Beyond free-porphyrin, porphyrin and phthalocyanine complexes can host a variety of strongly-correlated effects at their center that are useful for photochemistry and catalysis,\cite{ic_romelt,ic_romelt_2} among many other applications. Overall, and despite their great importance in photochemistry,\cite{hoshino_jacs,horvath_ccs,shinsuke_jpcb} spin manipulation,\cite{rawson_jpcb,cho_jacs} and biological applications,\cite{olafson_pnas,goff_jacs,ricchelli_jppbb} these strongly-correlated complexes remain very difficult for \textit{ab-initio} theory. The complexes are simply too large for most methods of high accuracy but too correlated to use a less expensive theory. Continued development of methods which address the challenge of strongly-correlated states embedded in a large host molecule is necessary to advance the field.

We have recently developed an active space theory called dynamical configuration interaction (DCI) for these challenging multiconfigurational systems.\cite{dvorak_prb,dvorak_prm} In an active space of strongly-correlated orbitals, we use full configuration interaction to treat static correlation. Additionally, the effects of surrounding degrees of freedom are included at a more efficient level of theory based on the many-body Green's function. The embedding construction interpolates between wave function and Green's function methods to parallel the physical problem of a central atom embedded in a porphyrin. The theory is efficient and accurate for small molecules,\cite{dvorak_prm} and its concept is well-suited to systems like porphyrin complexes. In this letter, we test DCI on porphyrin molecules for the first time, our largest systems studied to date. Our DCI calculations give excellent agreement with benchmark theory and experiment for low computational cost, encouraging further research to apply DCI to strongly-correlated porphyrin complexes.

Our spectroscopic perspective on static and dynamic correlation also serves as a bridge to discuss different methods. Identical arguments about static and dynamic correlation in optical spectra also apply to peaks in photoemission (PE) spectra describing particle addition or removal. In the language of the electron propagator, or single-particle Green's function (GF),\cite{fetter_quantum,cederbaum_domcke,ann_ortiz,ortiz_review} a sharp, dynamically correlated peak in the PE spectrum has a long quasiparticle lifetime. The appeal of propagator methods is that the spectral function (a good approximation to the PE spectrum) can be calculated in an effective single-particle picture.\cite{ortiz_ijqc,cederbaum_comp_phys} This efficient, effective single-particle picture is an important building block for DCI, hence our brief mention of PE spectra and its relation to theory. It is common in propagator methods applied to weakly-correlated systems to ignore the imaginary part of the self-energy altogether, equivalent to assuming an infinite quasiparticle lifetime. This long-lived quasiparticle picture works well for weakly-correlated systems and can be applied for relatively low cost. Perturbative corrections based on the $GW$ approximation\cite{pr_hedin,front_golze,wires_reining,jctc_knight} in Green's function theory, in particular, are good for this purpose.

In contrast, static correlation in the spectral function is better explained by considering the Lehmann amplitudes, or overlaps between the particle addition/removal state with any eigenstate.\cite{fetter_quantum} If many of these amplitudes are large and closely spaced in energy, the long-lived quasiparticle picture is not valid. It is trivial to describe this static correlation with the Lehmann amplitudes and compute the spectral function, \textit{if} one already has all the eigenstates of the system. In practice, this is an impossibly expensive task for real systems. The strong mixing of eigenstates in the spectral function, however, demonstrates that the common quasiparticle theories are inadequate for this regime of correlation. A more reliable choice for static correlation is to directly compute the eigenstates with a wave function (WF) method.

Wave function and Green's function theories are therefore complementary to each other. It is possible to exactly describe any interacting system with either choice. However, it is simpler to describe static correlation with the WF instead of the GF as long as the effective system size is kept small. Conversely, dynamic correlation is more efficiently described with GF theory, which scales better than WF theory, as long as low-energy channels which transfer spectral weight away from the quasiparticle solution do not exist or are removed. These properties suggest that a hybrid GF-WF embedding theory could be successful and efficient by treating static correlation with the WF and dynamic correlation at the quasiparticle level of theory.

The complementary nature of WF and GF theories is the motivation for the DCI quantum embedding theory \cite{dvorak_prm,dvorak_prb}. We present a brief summary of DCI here and refer the reader to Ref.~\onlinecite{dvorak_prb} for the complete theory. Based on the L\"owdin downfolding\cite{jmp_lowdin} of the many-body Hamiltonian ($\mathcal{H}$), the many-body Hilbert space is separated into two parts defined by projection operators $\mathcal{D}$ (strongly-correlated) and $\mathcal{R}$ (weakly-correlated). The full Hamiltonian is downfolded onto an energy-dependent effective Hamiltonian in the $\mathcal{D}$ space by the standard technique as
\begin{eqnarray}
M(E) &=& \left[ \mathcal{D} \mathcal{H} \mathcal{R} \right] \frac{1}{E - \mathcal{R} H \mathcal{R}} \left[ \mathcal{R} \mathcal{H} \mathcal{D} \right] \label{correction} \\
\mathcal{H}^{\mathrm{eff}}(E) &=& \mathcal{D} \mathcal{H} \mathcal{D} + M(E)  \\
\mathcal{H}^{\mathrm{eff}}(E) \ket{\psi} &=& E \ket{\psi} \; .  \label{downfolded}  
\end{eqnarray}

The concept of the DCI theory is to then rewrite the projected Hamiltonian $\mathcal{R} \mathcal{H} \mathcal{R}$ as correlated excitations ($\Omega^{\mathcal{R}}$) propagating above a correlated ground state ($E_0^{\mathcal{R}}$):
\begin{equation}
\mathcal{R} \mathcal{H} \mathcal{R} \rightarrow H^{\mathcal{R}} = E_0^{\mathcal{R}} + \Omega^{\mathcal{R}} \; . \label{qp_renormalization}
\end{equation}
Eq.~\ref{qp_renormalization} is the critical step where we flip the description from a WF theory on the LHS to a GF theory on the RHS. The quasiparticle renormalization of the $\mathcal{R}$ subspace in Eq.~\ref{qp_renormalization} is set up for the methodology of many-body Green's functions. Roughly speaking, bare particles contributing to any $\mathcal{R}$ excitation are replaced by their infinitely long-lived quasiparticle counterpart. Most importantly, calculations of $E_0^{\mathcal{R}}$ and $\Omega^{\mathcal{R}}$ can include \textit{only} intra-$\mathcal{R}$ correlation to avoid double-counting errors. The DCI theory is carefully constructed to satisfy two limits on the embedding to recover either a normal GF calculation ($\mathcal{R} \rightarrow \mathbb{I}$) or WF calculation ($\mathcal{D} \rightarrow \mathbb{I}$). Interpolating between these two limits, as we do for the actual embedding problem, gives a quasiparticle calculation entirely contained in the $\mathcal{R}$ subspace and leaves the overall theory double-counting free.

Once the $\mathcal{R}$ subspace excitation energies and ground state energy are known, $H^{\mathcal{R}}$ is inserted into the denominator of $M(E)$ in Eq.~\ref{correction} in place of $\mathcal{R} \mathcal{H} \mathcal{R}$. The hybridization between the two spaces is an exact matrix element of $\mathcal{D} \mathcal{H} \mathcal{R}$ evaluated with the Slater-Condon rules. To reduce the computational cost, we adopt a diagonal approximation to $\mathcal{H}^{\mathcal{R}}$.  Even in the diagonal $\mathcal{R}$ approximation, correlation in the $\mathcal{R}$ subspace is still treated at the quasiparticle level of theory. Our diagonal approximation is computationally efficient and allows us to embed the active space in a huge number of bath orbitals \textit{which are themselves correlated}. We adopt the very successful $GW$ approximation to the quasiparticle self-energy\cite{pr_hedin,hybertsen_louie_prl,front_golze,prb_37_godby,prl_56_godby,aryasetiawan_review,wires_reining}, which is a sum to infinite order of a certain type of diagram in perturbation theory. Furthermore, we extend the $GW$ approximation to a Bethe-Salpeter equation-like interaction between quasiparticles. As we will show later, the theory can be quantitatively compared to the diagonal, bare $\mathcal{R} \mathcal{H} \mathcal{R}$ Hamiltonian, and the effect of $\mathcal{R}$ correlation is a significant improvement in the results.

An important step in the DCI construction is to calculate the ground state energy $E_0^{\mathcal{R}}$. While $E_0^{\mathcal{R}}$ is difficult to formally define since the ground state exists primarily in $\mathcal{D}$, a sensible calculation is still possible with physical considerations and by enforcing the limits of the embedding. $E_0^{\mathcal{R}}$ has meaning as a ground state total energy and can be connected to the \textit{true} ground state energy as $\mathcal{R} \rightarrow \mathbb{I}$. Accordingly, $E_0^{\mathcal{R}}$ should be a size extensive quantity, as any total energy is.

As a methodological improvement over our past work, we calculate $E_0^{\mathcal{R}}$ with the constrained random phase approximation (cRPA).\cite{sasioglu_prb_83,werner_prb_91,biermann_prb_86} The cRPA is a major component of our overall approach and is discussed more in Supp. Info. The cRPA omits $\mathcal{R}$ screening channels from the polarizability when calculating either the partially screened Coulomb interaction ($W_{\mathcal{R}}$) or correlation energy.\cite{prb_aryasetiawan,prb_friedrich} To compute $E_0^{\mathcal{R}}$, we first calculate a correlation energy based on the cRPA, $C^{\mathrm{cRPA}}$. We add $C^{\mathrm{cRPA}}$ to the energy of the reference configuration as
\begin{equation}
E_0^{\mathcal{R}} = E^{\mathrm{ref}} + C^{\mathrm{cRPA}} \; .  \label{crpa_corr}
\end{equation}
Eq.~\ref{crpa_corr} correctly satisfies the limits of the embedding by recovering the full RPA total energy as $\mathcal{R} \rightarrow \mathbb{I}$ and zero correlation as $\mathcal{R} \rightarrow 0$. As the $\mathcal{R}$ subspace changes size, $E_0^{\mathcal{R}}$ in Eq.~\ref{crpa_corr} adjusts to match the size of the $\mathcal{R}$ subspace.

Just as the cRPA correlation energy adjusts as $\mathcal{R}$ changes size, the partially screened Coulomb interaction ($W_{\mathcal{R}}$) is also size extensive. The screened Coulomb interaction is a main ingredient of Green's function theory. Physically, it represents the response of the microscopic degrees of freedom to screen and weaken the interaction between two charges. In our case, $W_{\mathcal{R}}$ is an ingredient for the self-energy assigned to $\mathcal{R}$ subspace quasiparticles and also describes inter-quasiparticle interactions. As more configurations are added to $\mathcal{R}$, quasiparticles are further screened, correctly reflecting the change to correlation in the $\mathcal{R}$ subspace. $W_{\mathcal{R}}$ is a favorite tool of strongly-correlated physics,\cite{prb_aryasetiawan,prb_friedrich} but it has gained attention recently in our own and other CI-based embedding schemes.\cite{bockstedte_npjqm_3,arxiv_ma} As we show below, these two effects ($C^{\mathrm{cRPA}}$ and $W_{\mathcal{R}}$) appear to recover a significant amount of correlation that is absent from the bare $\mathcal{RHR}$ Hamiltonian.

\begin{figure}[H]
\begin{centering}
\includegraphics[width=\columnwidth]{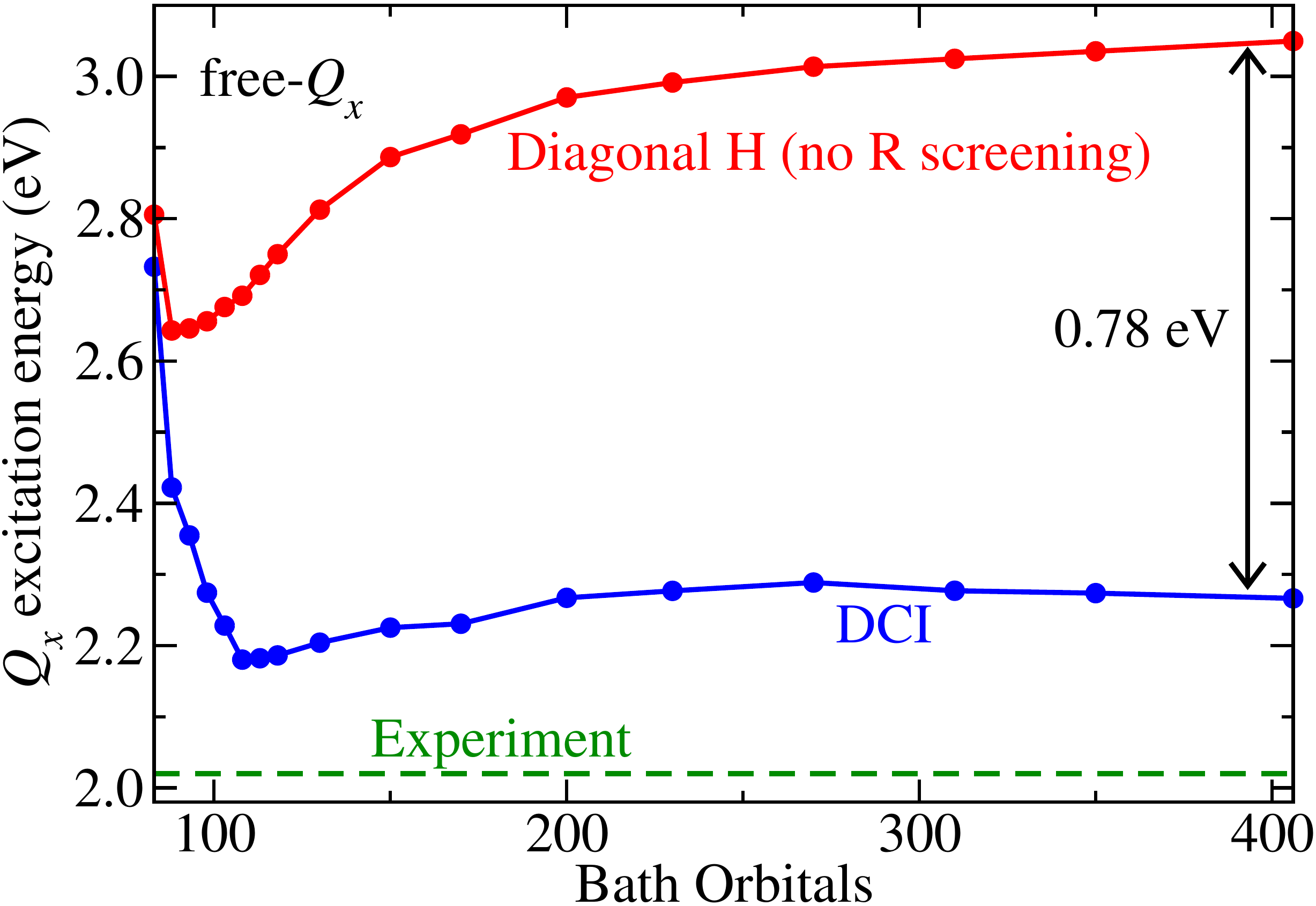}
\includegraphics[width=\columnwidth]{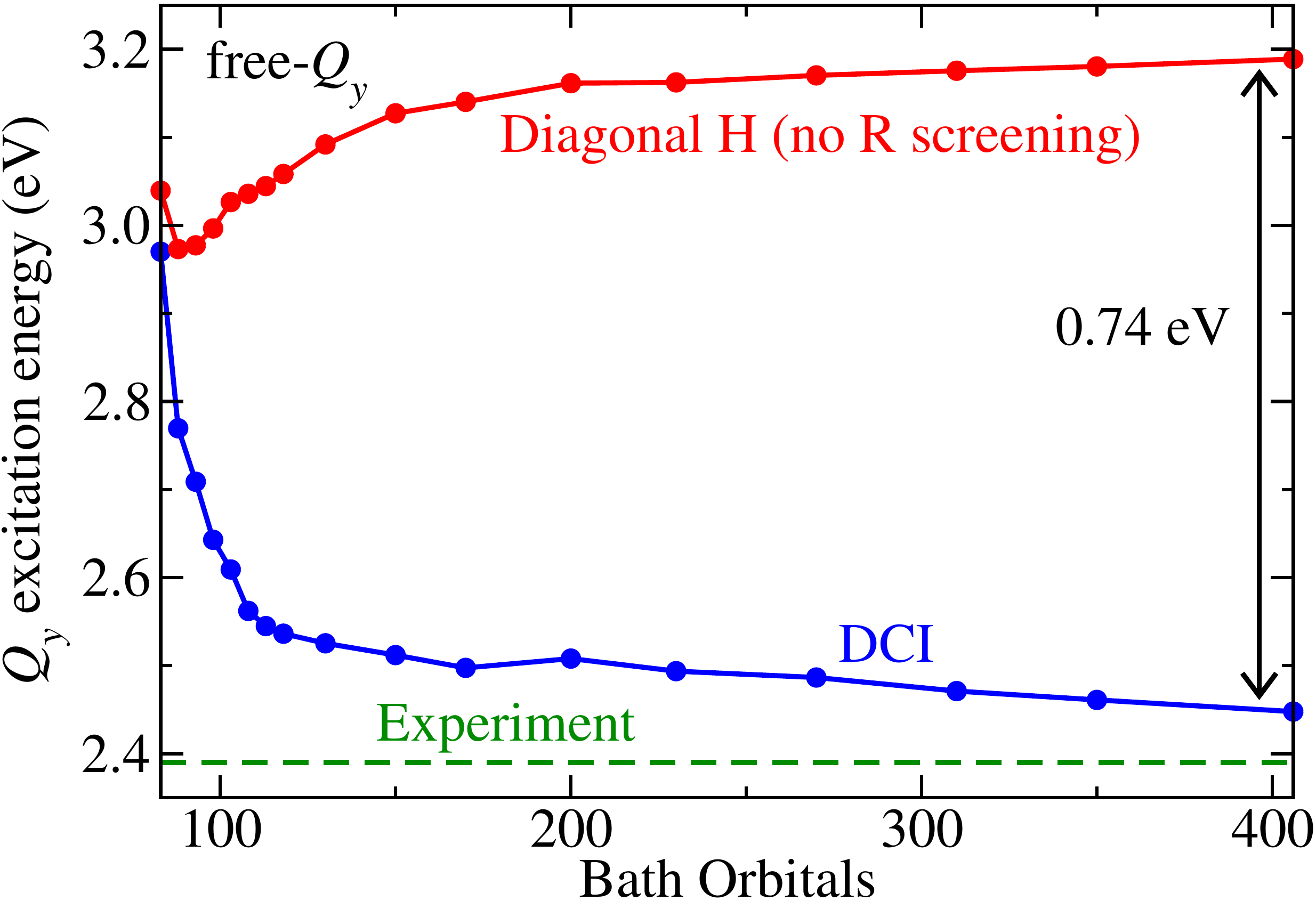}
\includegraphics[width=\columnwidth]{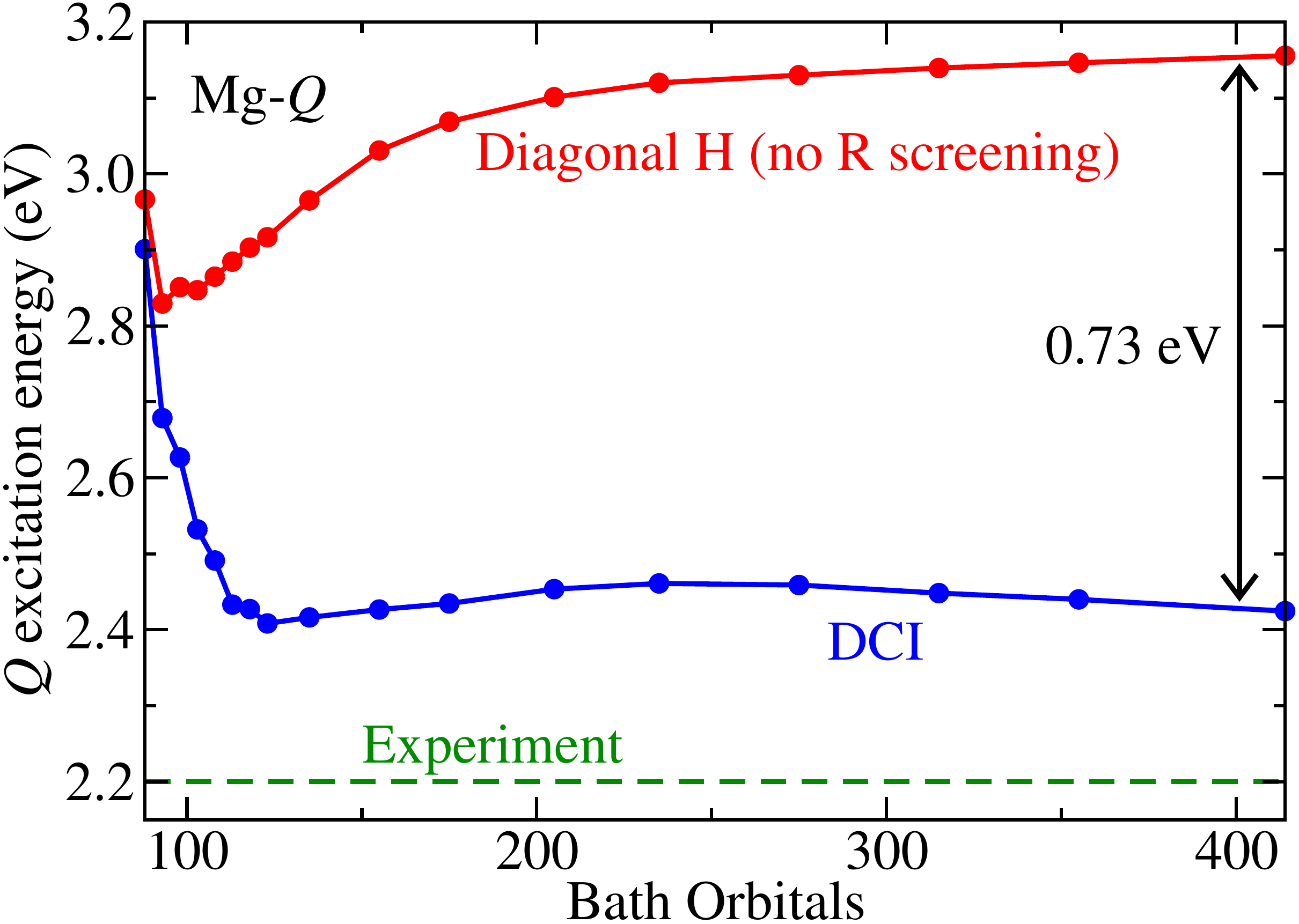}
\caption{Convergence of virtual orbitals for the $Q_x$, $Q_y$, and $Q$ excitations of free- and Mg-porphyrin with DCI(4,4). The HOMO of free- (Mg-) porphyrin is number 81 (86). Calculations are based on FHI-AIMS and use the cc-pVDZ Dunning basis set.\cite{jcp_dunning} \label{porphyrin_data}}
\end{centering}
\end{figure}

The DCI computational procedure first generates $\mathcal{D}$ as in a complete active space (CAS) theory for a set of strongly-correlated orbitals. These CAS configurations, generated as full configuration interaction of the single-particle active space, are designated reference configurations and placed in $\mathcal{D}$. Next, we sample the $\mathcal{R}$ space in a way analogous to multireference CI with single and double excitations (MR-CISD). Treating each $\mathcal{D}$ configuration as a reference, we generate a local set of bath configurations as all possible single and double excitations from that reference. For a CAS treatment of four electrons in four levels, denoted (4,4), the highest excited reference configuration is quadruple, and the highest excited bath configuration is $6\times$. With this procedure, DCI retains the best features of MR-CISD, like an exact treatment of static correlation and a balanced treatment of ground and excited states, while lowering the computational cost with the diagonal $\mathcal{R}$ approximation. Even at a lower cost than MR-CISD, however, the theory still includes $\mathcal{R}$ correlation at a very successful level of theory.

Porphyrins are an excellent test of any active space theory because of the apparent importance of ligand correlation for quantitative accuracy of the $Q_x$ and $Q_y$ excitations. There is some discussion in the literature that the $Q_y$ transition of free-porphyrin is more correlated with the surrounding set of orbitals than $Q_x$.\cite{ijqc_yamamoto,cpl_merchan,manni_jctc} Most theories struggle with the sheer number of orbitals which must be correlated to quantitatively test this idea. However, correlating a large number of orbitals for low computational cost is the strength of DCI. We are able to include dynamic correlation with our MR-CISD-like basis generated from a bath of $\sim$400 orbitals (the full set in our chosen atomic basis). Overall, porphyrins are a demanding benchmark to test how well dynamic correlation outside the minimal active space is included in the DCI calculation $-$ a test that can be difficult for multiconfigurational-SCF theories, even with a second-order perturbative (PT2) correction.

We perform DCI(4,4) calculations for free- and Mg-porphyrin. Our calculations of the vertical $Q_x$ and $Q_y$ excitation energies embed the four Gouterman orbitals in the full set of molecular orbitals. Our calculations are fully \textit{ab-initio} and parameter free. For comparison, we calculate the same excitation energies using diagonal matrix elements of the bare $\mathcal{RHR}$ Hamiltonian, which we denote ``Diagonal H" in Fig.~\ref{porphyrin_data}. The Diagonal H approximation neglects correlation in $\mathcal{R}$. By comparing DCI to Diagonal H, we can quantify the effect of screening and correlation in the $\mathcal{R}$ subspace on the $Q$ excitation energies. The convergence of DCI and Diagonal H calculations as a function of bath orbitals is shown in Fig.~\ref{porphyrin_data}. For zero bath orbitals on the scale of Fig.~\ref{porphyrin_data}, the calculation only includes static correlation in the four-level Gouterman subspace. In that case of zero bath orbitals, not shown in Fig.~\ref{porphyrin_data} but listed in Table~\ref{porphyrin_table}, DCI and Diagonal H are identical. Indeed, for a low number of orbitals on the horizontal axis of the Fig.~\ref{porphyrin_data}, the DCI and Diagonal H curves approach each other. Bath orbitals are then added to the calculation until the active space is embedded in the full set of molecular orbitals.

For all three excitations considered here, the difference between DCI and Diagonal H is a substantial $\sim$0.75 eV. $\mathcal{R}$ correlation noticeably lowers the excitation energies and improves agreement with experiment. Notably, excitation energies with the Diagonal H approximation continue to increase as the number of bath orbitals is converged. Diagonal H appears to miss a great deal of additional correlation that would lower the excitation energy towards the experimental value as the basis grows. This error is expected since there is no mixing, or correlation, of configurations within the $\mathcal{R}$ space, and the overall basis is truncated to a finite excitation level. On the other hand, the cRPA correlation energy included in DCI immediately shifts the effective ground state in $\mathcal{R}$ lower than that of Diagonal H. Quasiparticle self-energies used in DCI also mix $\mathcal{R}$ configurations, further adding correlation to the calculation. The convergence of the DCI calculations appears flat or slightly downward in comparison to Diagonal H, indicating that much of the correlation missed by Diagonal H is recovered with DCI. 

\begin{table}[htb]
\begin{tabular}{r|c|c}
          & $Q_x$ & $Q_y$ \\  \hline
Exp.\cite{jmp_edwards}      & 2.02  & 2.39  \\
DCI(4,4)$\;\;\;$       & 2.27  & 2.45  \\
MC-CASSCF(26,24)\cite{manni_jctc} & 2.02  & 3.24  \\
RASSCF(26,24)\cite{manni_jctc}    & 2.55  & 3.38  \\
RASPT2(26,24)\cite{manni_jctc}    & 2.54  & 2.81  \\
Frozen core (zero bath) $\;\;$          & 3.48  & 3.67
\caption{Vertical $Q_x$ and $Q_y$ excitation energies (eV) of free-porphyrin computed with DCI and reference data in similar (not identical) basis sets. MC-CASSCF refers to the stochastic CASSCF method of Ref.~\cite{manni_jctc} \label{porphyrin_table}}
\end{tabular}
\end{table}

\begin{table}[htb]
\begin{tabular}{r|c}
              & $Q$  \\  \hline
Exp.\cite{cpl_starukhin}          & 2.20 \\
DCI(4,4)$\;\;$          & 2.42 \\
MRPT\cite{jpca_hashimoto}          & 2.00 \\
MC-CASSCF(26,24)\cite{manni_jctc}     & 2.50 \\
CASSCF(15,18)\cite{cpl_serrano} & 3.06 \\
CASPT2(15,18)\cite{cpl_serrano} & 1.78 \\
Frozen core (zero bath) $\; \;$   &   3.60 
\caption{Vertical $Q$ excitation energy (eV) of Mg-porphyrin computed with DCI and reference data in similar (not identical) basis sets. By symmetry, $Q_x$ and $Q_y$ merge into the degenerate $Q$ in Mg-porphyrin. MC-CASSCF refers to the stochastic CASSCF method of Ref.~\cite{manni_jctc}. MRPT denotes multireference perturbation theory and CASPT2 denotes a second-order perturbation correction applied to a complete active space self-consistent-field wave function. \label{mg_porphyrin_table}}
\end{tabular}
\end{table}

Values for $Q$, $Q_x$ and $Q_y$ from high level theoretical reference data (in similar basis sets to our own) vary, as shown in Tables~\ref{porphyrin_table} and \ref{mg_porphyrin_table}. One of the conclusions from the impressive stochastic complete active space self-consistent-field (CASSCF) study of Ref.~\onlinecite{manni_jctc} is that dynamic correlation is more important in free-porphyrin for the $Q_y$ excitation than $Q_x$. This is indicated by the significant shift of $Q_y$ from restricted active space self-consistent-field theory (RASSCF) to RASPT2 in Table~\ref{porphyrin_table} that is induced by additional dynamic correlation in the RASPT2 calculation (RASPT2 is RASSCF plus an additional second-order perturbation theory). Our DCI calculation of the $Q_y$ excitation energy converges slower than $Q_x$, corroborating this picture. The convergence curve of $Q_y$ does not show a deep local minimum as $Q_x$ does. Roughly speaking, a calculation at the local minimum for $Q_x$ could partially compensate the missing bath orbitals. $Q_y$, however, decreases monotonically, and no such compensation is possible. The slower convergence of $Q_y$ supports the idea that $Q_y$ is more sensitive to $\sigma$ orbital ligand correlation than $Q_x$. The effect of restricting certain excitation levels in RASSCF compared to CASSCF is also noticeable. A direct comparison of excitation levels between DCI and the multiconfigurational-SCF methods is not possible.

The situation for Mg-porphyrin is similar. The difference in $Q$ excitation energy between CASSCF and CASPT2 again indicates the importance of dynamic correlation outside the active space. Efficiently and accurately adding this dynamic correlation to CAS theories is a challenge, but this is where DCI excels. The sensitivity to the choice of active space is also evident by comparing MC-CASSCF(26,24) to CASSCF(15,18). It is clearly important to correlate as many orbitals as possible for quantitative accuracy. Across these three excitation energies, the overall performance of DCI with the minimal active space is competitive with these more expensive methods. The error between DCI and experiment is $\sim$0.2 eV or less, with DCI overestimating experiment in all cases.

A standard approach to computing optical excitations in condensed matter physics and of growing relevance in quantum chemistry\cite{jacquemin_jctc,blase_csr,jacquemin_jpcl,liu_jcp} is the Bethe-Salpeter equation (BSE).\cite{Salpeter/etal:1951,onida_prl_75,albrecht_prl_80,hanke_prl_43,strinati:1984,strinati:1982,gf_review} The BSE describes the correlated propagation of an electron and hole via the two-particle correlation function. Combined with the $GW$ approximation, the $GW$-BSE method with a standard set of assumptions has become the state-of-the-art tool for computing optical properties of dynamically correlated solid-state systems. Our $\mathcal{R}$ space treatment of quasiparticle excitations is deliberately $GW$-BSE-like to capitalize on this success.

For this reason, we present a careful comparison between DCI and $GW$-BSE. DCI is constructed to recover the successful description of dynamic correlation that $GW$-BSE is known to provide.\cite{jacquemin_jpcl,blase_csr,onida_prl_75,albrecht_prl_80} To test this idea, we perform our own $GW$-BSE calculations in FHI-AIMS\cite{liu_jcp} with the exact same basis set and numerical settings as our DCI calculations. Our results are tabulated in Table~\ref{gw-bse_table}. The numbers that can be directly compared to each other are DCI and $G_0W_0$-BSE at the same starting point, which is Hartree-Fock (denoted $G_0W_0$-BSE@HF). These two calculations are remarkably close to each other and in close agreement with experiment. We point out that, numerically speaking, it is accidental that $G_0W_0$-BSE@HF and DCI are so close $-$ the calculations themselves are substantially different. The good agreement indicates that DCI does recover dynamic correlation at a level similar to $GW$-BSE, as we intended with the construction of the theory. In this case, and based on the agreement between both methods with experiment, it appears that dynamic correlation dominates in these porphyrins. This picture is consistent with the variation in results from multiconfigurational-CAS theories and their sensitivity to active space: it is important to dynamically correlate all orbitals.

Although it is not necessarily indicated in the head-to-head comparison between DCI and $GW$-BSE for these porphyrins, it is important to point out that DCI has the \textit{additional} advantage of trivially handling static correlation and multiple excitations in the active space. The ideal theory should be able to handle \textit{both} types of correlation, and DCI shows promise in this regard. The impressive description of dynamic correlation with DCI encourages future studies of more challenging, statically correlated systems like porphyrin complexes for which purely dynamically correlated methods (like $GW$-BSE) fail and multiconfigurational-SCF methods would require too large an active space. We find worse agreement between experiment and $GW$-BSE with a density functional starting point. The difference between our $GW$-BSE results and previous work\cite{jcp_palummo} based on a density functional starting point could be due to different basis sets and implementations.

\begin{table}[]
\begin{tabular}{r|c|c|c}
          & $Q_x$ & $Q_y$ & $Q$ \\  \hline
Exp.\cite{jmp_edwards,cpl_starukhin}      & 2.02  & 2.39 & 2.20 \\
DCI $\;\;\;$    &  2.27 & 2.45 & 2.42    \\
$G_0W_0$-BSE@HF$^*$   $\;$   & 2.27  &  2.47  &  2.41  \\
$G_0W_0$-BSE@LDA$^*$ $\;$    & 1.62  &  1.87  &  1.82  \\
$G_0W_0$-BSE@LDA \cite{jcp_palummo} & 1.98 & 2.30 & --  \\
$G_0W_0$-BSE@PBE$^*$  $\;$   & 1.62  &  1.88  &  1.82  
\caption{Vertical $Q_x$ and $Q_y$ excitation energies (eV) of free-porphyrin and Mg-porphyrin ($Q$) computed with $G_0W_0$-BSE at the indicated starting point. Calculations with $^*$ are our own calculations with FHI-AIMS in the same cc-pVDZ basis and numerical settings as DCI calculations. \label{gw-bse_table}}
\end{tabular}
\end{table}

In conclusion, our dynamical configuration interaction calculations of $Q$ excitations in free- and Mg-porphyrin compare very well with experiment and benchmark theoretical results. DCI shares certain features of MR-CISD but at lower computational cost. These conclusions make DCI an attractive fully \textit{ab-initio} embedding theory for large chemical systems going forward, including strongly-correlated porphyrin complexes. The theory's greatest strengths are its ability correlate a large number of orbitals for low cost and its balance of static and dynamic correlation. Our future work is aimed at systematically benchmarking DCI for small organic molecules, tackling $d$-electrons, and studying more complicated porphyrin complexes.

\section*{Computational Details}
\label{sec:computational}
Our calculations are based on the FHI-AIMS electronic structure package.\cite{cpc_blum,njp_ren,njp_ihrig} FHI-AIMS is based on numeric atomic orbitals and uses the resolution of the identity (RI) technique for four-center integrals. We use the Dunning\cite{jcp_dunning} cc-pVDZ basis sets throughout and employ the analytic continuation technique for the $GW$ self-energies. Self-energies are evaluated in the $G_0W_0$ approximation. Our calculations always use a restricted Hartree-Fock starting point.

When solving the effective Hamiltonian, each eigenvalue is found self-consistently by iterating the effective Hamiltonian. We do not linearize any portion of the calculation or make further approximations beyond those discussed in this work and Ref.~\onlinecite{dvorak_prb}.

Molecular geometries are taken from Refs.~\onlinecite{manni_jctc} and \onlinecite{jcp_nagashima}. To support open science, \cite{Himanen/Geurts/Foster/Rinke:2019} we made the results of all relevant calculations available on the Novel Materials Discovery (NOMAD) repository.\cite{NOMAD} 

%
\begin{acknowledgement}
We acknowledge CSC -- IT Center for Science, Finland, and the Aalto Science-IT project for generous computational resources. This work was supported by the Academy of Finland through grant no.~316347.
\end{acknowledgement}
 
\begin{suppinfo}
Supporting Information available online.
\end{suppinfo}


\bibliography{main}

\begin{mcitethebibliography}{63}
\bibitem[NOM()]{NOMAD}
To avoid the generation of multiple DOIs, submission to NOMAD repository will
  follow inclusion of reviewer's comments.\relax
\mciteBstWouldAddEndPunctfalse
\mciteSetBstMidEndSepPunct{\mcitedefaultmidpunct}
{}{\mcitedefaultseppunct}\relax
\EndOfBibitem
\end{mcitethebibliography}

\end{document}